\documentstyle[12pt]{article}
\newcommand{\be}{\begin{equation}}
\newcommand{\ee}{\end{equation}}
\newcommand{\ba}{\begin{eqnarray}}
\newcommand{\ea}{\end{eqnarray}}

\begin{document}
\renewcommand{\baselinestretch}{1.4}
\small\normalsize
\renewcommand{\thesection}{\arabic{section}.}
\renewcommand{\thefootnote}{\fnsymbol{footnote}}
\language0

\vspace*{1cm}

\begin{center}

{\Large \bf Cluster dynamics for first-order phase transitions
            in the Potts model}

\vspace*{0.8cm}

{\bf Werner Kerler and Andreas Weber}

\vspace*{0.3cm}

{\sl Fachbereich Physik, Universit\"at Marburg, D-35032 Marburg, Germany}

\end{center}

\vspace*{1.5cm}

\begin{abstract}
An algorithm for Monte Carlo simulations is proposed in which the parameter
controlling the strength of the transition becomes a dynamical variable and
in which efficient transitions are achieved by cluster steps. Numerical results
for the Potts model demonstrate the advantages of the method.
\end{abstract}

\newpage

To overcome slowing down is crucial in computer simulations of phase
transitions. For second-order phase transitions in a number of systems critical
slowing down can be drastically reduced by cluster algorithms as introduced by
Swendsen and Wang \cite{sw}. There the acceleration is achieved by the nonlocal
update of larger structures.

In the case of first-order transitions one encounters a different problem. On
finite lattices one has an exponentially fast suppression of the tunneling
between metastable states of the system with increasing lattice size. To reduce
this type of slowing down the multicanonical Monte Carlo algorithm has been
proposed \cite{bn,bn2} which uses enhancement of the suppressed configurations.
The canonical distribution is then obtained by an appropriate reweighting
\cite{fs}.

The application of cluster algorithms to first-order transitions leads only to
little improvement \cite{bip,ja1}. The obvious reason for this is that clusters
can lead to some acceleration related to extended structures at not too large
$q$, however, not with respect to the tunneling between order and disorder.

To overcome slowing down in systems with a rough free-energy landscape the
method of simulated tempering \cite{mp} has been proposed and applied to the
random-field Ising model. In this method temperature values are related to a
dynamical variable and the fact is utilized that at lower $\beta$ the
free-energy barriers are lower.

In the present paper we propose to make the parameter which controls the
strength of a first order transition a dynamical variable. This allows to
proceed from a state at large strength to ones at lower strength where the
transitions become easier. If in particular there is a range where the
transition gets of second order, in this way the tunneling and the related
suppression can be completely circumvented. It is crucial then to avoid also
the critical slowing down related to extended structures. This is achieved by
combining the procedure with cluster updates.

We demonstrate in the case of the Potts model \cite{po} in two
dimensions, using a square lattice with periodic boundary conditions, that this
general scheme works very efficiently. The parameter which has to become
dynamical in this case is $q$, the number of the degrees of freedom of the
spins.

Figure 1 shows the distribution $P(E,q)$ for energies and values $q$ which is
obtained for $\beta_L$ (defined below). It is useful to illustrate the
principle
of our approach. Instead of tunneling across the valley we allow travelling
along the peaks.

Starting at an ordered state at large strength of the transition the cluster
updates will mainly induce transitions between ordered states. Travelling
down to lower strength they will increasingly break up larger structures.
{}From low strength then an easy way is open to disorder at large strength.
Further, also all the steps between different values of the strength variable
will contribute to decorrelation.

Before describing our update scheme it appears appropriate to formulate the
general principle of making a parameter dynamical. A probability distribution
$\mu_q(\sigma)$ of spin configurations $\{\sigma\}$ which depends on a
parameter
$q$ can be reinterpreted as the conditioned probability $\hat{\mu}(q;\sigma)$
for getting a configuration $\{\sigma\}$ given a value of $q$. Then prescribing
a distribution $\bar{\mu}(q)$ one obtains the joint distribution
$\mu(q,\sigma)=
\bar{\mu}(q)\hat{\mu}(q;\sigma)$ for which the updates are to be performed.

Our update scheme has three steps. The first one generates bond configurations
$\{n\}$ according to Swendsen and Wang \cite{sw} and is described here by the
conditioned probability $A(\sigma,q;n)$. The second step is a Metropolis
update for the marginal distribution $\tilde{\mu}(q,n)=\sum_{\{\sigma\}}
\mu(q,\sigma)A(\sigma,q;n)$ realizing the conditioned probability $T(n,q;q')$
for arriving at a new value $q'$. In the third step the new spins are obtained
with the conditioned probability $\tilde{A}(n,q';\sigma')=\mu(q',\sigma')
A(\sigma',q';n)/\tilde{\mu}(q',n)$ which corresponds to the special choice of
Swendsen and Wang where the new cluster spins are independent of the old ones.
The generalized transition matrix thus becomes
\be
\label{w}
W(\sigma,q;\sigma',q')=\sum_{\{n\}}A(\sigma,q;n)T(n,q;q')\tilde{A}(n,q';\sigma')
\quad .
\ee
It leaves $\mu$ invariant if $T$ leaves $\tilde{\mu}$ invariant, and it
satisfies detailed balance with $\mu$ if $T$ does so with $\tilde{\mu}$, which
follows immediately from the definitions.

In our update scheme cluster generation in addition to reducing critical
slowing
down also has the important virtue to allow technically simple steps between
different $q$-values. The reason for this is that spins are not involved and
the
dependences on the bond configurations are simple in $\tilde{\mu}(q,n)$. Using
$\hat{\mu}(q;\sigma) \equiv \mu_q(\sigma)=Z^{-1}e^{-\beta H}$ with
$H=-\sum_{\langle ij\rangle} \delta_{\sigma_i,\sigma_j}$ and
$Z=\sum_{\{\sigma\}}e^{-\beta H}$, one gets its explicit form
\be
\label{tmu}
\tilde{\mu}(q,n)=\bar{\mu}(q)Z^{-1}(q)(e^{\beta(q)}-1)^{N_b(n)}q^{N_c(n)}
\ee
where $N_b$ is the number of bonds and $N_c$ the number of clusters of the
bond configuration $\{n\}$.

The simulations have been performed for two sets of $q$-values, $q=4,
\ldots,7$ and $q=2,\ldots,10$ . Detailed balance for the Metropolis steps
described by $T(n,q;q')$ has been guaranteed using the symmetric proposal
matrix
$T_p(q;q')=\frac{1}{2}(\delta_{q+1,q'}+ \delta_{q,q'+1}+\delta_{q,q_{min}}
\delta_{q',q_{min}}+\delta_{q,q_{max}} \delta_{q',q_{max}}$) and the acceptance
matrix min$(1,\tilde{\mu}(n,q')/ \tilde{\mu}(n,q))$.

The distribution $\bar{\mu}(q)$ has been choosen to be (approximately) constant
within the sets of $q$-values considered. In this context also the
$q$-dependence of $Z^{-1}(q)$ in (\ref{tmu}) is to be known. By requiring the
number of sweeps to be the same for each of the $q$-values in the respective
set
it has been determined numerically in an iterative way. Table 1 shows the
relative values of $Z^{-1}(q)\bar{\mu}(q)$ obtained for the set $q=4,\ldots,7$
requiring constancy of $\bar{\mu}(q)$ better than 4\% .

For the values of $\beta(q)$ two choices have been considered. The first one,
denoted by $\beta_L(q)$ here corresponds to equal height of the two peaks in
the
distribution of energies for $q>4$ or to the maximum of the specific heat which
is used for $q\le 4$. Table 2 gives the values of $\beta_L(q)$ used for the set
$q=4,\ldots,7$. The second $\beta$-value considered is the infinite-volume
critical value $\beta_c= $ln$(\sqrt{q}+1)$.

The observables measured for all $q$ are the energy $E$ and the order
parameter $M=(q\mbox{ max}_{(a)}(N_a)-1)/(q-1)$ with $N_a=V^{-1}\sum_i\delta_
{\sigma_i a}$ . For both of them in each case also exponential and integrated
autocorrelation times have been determined. Further, for comparison also
simulations for other algorithms have been performed and in addition tunneling
times determined. The statistics collected has been in the range of $3\times
10^6$ to $2\times 10^7$ sweeps for each $q$ and all algorithms investigated.

There are results on other algorithms in literature for $q=7$ \cite{bil,jbk,ru}
and for $q=10$ \cite{bn2,bip,bil}. We performed simulations for the heat bath
algorithm (H), the method of Swendsen and Wang (SW), and the multicanonical
heath-bath algorithm (MH) ourselves to get autocorrelation times for a
precise comparison (rather than to have to rely on tunneling times only
available in most cases). Because the Metropolis algorithm has been found
\cite{bip} to be rather slow it has not been considered here. SW turns out to
be slightly superior to H if one accounts for CPU times, too.

Tunneling times \cite{bip,jbk} are only useful at larger $q$ where they get
dominant in the autocorrelation times. We observe that they depend on the
particular definition and differ from the autocorrelation times, which for MH
and DQ persists even at larger $L$.

Table 3 gives the average stay times for each $q$ (the average number of
sweeps spent at a particular $q$ before leaving it) obtained for $\beta_L$ and
the set $q=4,\ldots,7$ . It is seen that the travelling goes relatively fast.
For $\beta_c$ and for the set $q=2,\ldots,10$ the stay times are very similar.

In Table 4 the exponential autocorrelation times for $\beta_L$ and the set
$q=4,\ldots,7$ are shown. The errors are statistical ones. In the cases of
$\beta_c$ and of the set $q=2,\ldots,10$ the results are very similar. The
results obtained for the order parameter within errors agree well with those
for E (listed in the Table). Thus the measurement of both observables provides
a valuable check.

Figures 2 and 3 give the exponential autocorrelation times for $q=7$ and
$q=10$, respectively, which we obtained for the heat bath algorithm (H), the
method of Swendsen and Wang (SW), the multicanonical heath-bath algorithm (MH),
and our approach with dynamical $q$ (DQ) for the sets $q=4,\ldots,7$ and
$q=2,\ldots,10$ . The time scale is in sweeps of the respective algorithm in
each case.

For a comparison firstly the CPU times related to these scales are to be
considered. Because the fraction of time for the Metropolis step is small (of
the order of 4\%) as compared to that of the cluster steps in DQ, these scales
for SW and DQ are about the same. For H and MH the underlying sweeps are
about a factor 2 slower.

To compare secondly one has to discuss that in DQ time is also spent at the
other values of $q$. With the extreme view that only one of the $q$-values is
of interest -- and all other results obtained at no extra cost are ignored --
one could multiply by 4 and by 9 in the cases of the sets $q=4,\ldots,7$
and $q=2,\ldots,10$ , respectively (noting that according to the choice of
$\bar{\mu}(q)$ equal times are spent at different $q$). Even then  one gets an
improvement as compared to MH of up to factors 7 and 3 for $q=7$ and $q=10$,
respectively, in the range considered \cite{add}. With the more realistic view
that all results are of interest the gain is more than an order of magnitude.

For $q=7$ by MH as compared to H there is only a modest gain \cite{jbk}
of about a factor of 2. Recently by a multicanonical demon algorithm \cite{ru}
some improvement upon MH has been achieved on larger lattices. For DQ on the
set
$q=4,\ldots,7$ we find an improvement with respect to both of these
multicanonical algorithms which even after a multiplication by 4 as mentioned
above is still about a factor 6 . Thus by DQ in any case one improves more than
an order of magnitude as compared to the conventional case.

For $q=10$ MH gets more efficient, gaining more than an order of magnitude in
the range considered here as compared to H. However, DQ remains still superior
even after the multiplication by 9 mentioned above. In addition, the
multiplication factor can be reduced by using the set $q=4,\ldots,10$ instead
of the set $q=2,\ldots,10$. This is suggested by the stay times indicating that
it is not worthwhile to go down to $q=2$. This feature is also apparent from
the
comparison of the results for DQ for $q=7$ obtained on the sets $q=4,\ldots,7$
and $q=2,\ldots,10$ where the advantage of not going down amounts to a factor
of about 2.

In this context it should be noted that there are a number of possibilities for
further improvements. Some have already been briefly tested with positive
results. For example, instead of $q$-steps by one one can use steps by two. To
accelerate certain transitions one can make $\beta$ dynamical, too. Instead of
keeping $\bar{\mu}(q)$ constant it may be adjusted to get more favorable stay
times. Modification of the Metropolis step offers various optimizations.

For $q=4$ and 3, where the phase transition gets of second order, comparing
with
integrated autocorrelation times obtained for SW \cite{ls} one observes still
reductions by factors of about 4 and 3, respectively. For $q=2$ comparing with
exponential autocorrelation times for SW \cite{ke} one gets a corresponding
factor of about 1.4 . To these reductions the additional tunneling possible at
not too large L contributes which does not persist at very large L.

{}From Figures 2 and 3 it is seen that for DQ there is some small
exponential-like
contribution present in the autocorrelation times. The obvious reason for this
is that at not too large $L$ there are still sizable additional contributions
from tunneling which die out at larger $L$. At very large L only travelling
along the peaks is possible and the true asymptotic behavior occurs.

At the upper end of the range considered for DQ the increase of autocorrelation
times is smaller than for MH which gives first hints about the ultimate
asymptotic behavior. However, to conclude about the ultimate behavior much
larger lattices are needed and it is to be noted that then details as, for
example, the appropriate adjustment of $\bar{\mu}(q)$ become important. From
the practical point of view on the lattices which can be reached there remains
in any case substantial gain for DQ.

The proposed concept has been realized here with the simplest choices of
details
in order to check its working in practice. Clearly the next task is to optimize
these choices. In any case the present results demonstrate that the concept
works surprisingly well such that it deserves further development and also
application to other systems.

\vspace{1cm}

One of us (W.K.) wishes to thank Claudio Rebbi and the Physics Department
of Boston University for their kind hospitality during a sabbatical leave.
This work has been supported in part by the Deutsche Forschungsgemeinschaft
through grants Ke 250/7-1 and 250/9-1. The computations have been done on the
SNI 400/40 of the Universities of Hessen at Darmstadt and on the Convex C230
of Marburg University.

\newpage

\renewcommand{\baselinestretch}{1.5}

\newpage

\renewcommand{\baselinestretch}{1.4}
\small\normalsize

\begin{center}

{\bf Table 1}

  Values of $(Z^{-1}(q)\bar{\mu}(q))/(Z^{-1}(7)\bar{\mu}(7))$ .

  \begin{center}
  \begin{tabular}{|c|c|c|c|}
  \hline
  $L$  & $q=6$ & $q=5$ & $q=4$ \\
  \hline
  12&$7.23\times10^{-1}$&$5.43\times10^{-1}$&
  $5.346\times10^{-1}$\\
  16&$4.748\times10^{-1}$&$2.629\times10^{-1}$&$
  1.925\times10^{-1}$\\
  24&$1.726\times10^{-1}$&$3.669\times10^{-2}$&$
  1.677\times10^{-2}$\\
  34&$2.857\times10^{-2}$&$1.18\times10^{-3}$&$
  1.95\times10^{-4}$\\
  50&$5.17\times10^{-4}$&$5.38\times10^{-7}$&
  $1.52\times10^{-8}$\\
  \hline
  \end{tabular}

\vspace{2cm}

{\bf Table 2}

Numerical values for $\beta_L$ .

  \begin{tabular}{|c|c|c|c|c|}
  \hline
  $L$ & $q=7$ & $q=6$ & $q=5$ & $q=4$ \\
  \hline
  12&1.2725&1.2158&1.1495&1.0708\\
  16&1.2806&1.2238&1.1582&1.0792\\
  24&1.2872&1.2309&1.1656&1.0879\\
  34&1.29005&1.23407&1.16915&1.0918\\
  50&1.29178&1.23607&1.17145&1.0947\\
  \hline
  \end{tabular}

\newpage

{\bf Table 3}

Average stay times for $\beta_L$ (with errors smaller than 1\%).

  \begin{tabular}{|c|c|c|c|c|}
  \hline
  $L$ &  $q=7$ & $q=6$ & $q=5$ & $q=4$ \\
  \hline
  12&2.48&1.27&1.32&2.70\\
  16&2.70&1.39&1.47&3.08\\
  24&3.22&1.69&1.88&4.02\\
  34&4.19&2.24&2.66&5.85\\
  50&6.84&3.81&5.04&12.8\\
  \hline
  \end{tabular}

\vspace{2cm}

{\bf Table 4}

Exponential autocorrelation times for $\beta_L$ .

  \begin{tabular}{|c|c|c|c|c|}
  \hline
  $L$ & $q=7$ & $q=6$ & $q=5$ & $q=4$ \\
  \hline
  12&15.1(2)&11.5(1)&9.04(6)&8.18(6)\\
  16&21.2(4)&15.8(2)&12.5(1)&10.9(1)\\
  24&36.8(6)&26.7(3)&20.4(2)&16.6(2)\\
  34&64.0(6)&44.7(5)&31.8(3)&24.1(3)\\
  50&126(2)&79(1)&50(1)&37.0(4)\\
  \hline
  \end{tabular}

\end{center}

\newpage
\renewcommand{\baselinestretch}{1.7}
\small\normalsize

{\large \bf Figure Captions}

\begin{tabular}{rl}
FIG. 1. & Distribution $P(E,q)$ for $\beta_L$ (shown for $L=34$).\\
FIG. 2. & Comparison of exponential autocorrelation times for $q=7$ and\\
        & the algorithms H, MH, DQ on set $q=4,\ldots,7$, and DQ on set\\
        & $q=2,\ldots,10$.\\
FIG. 3. & Comparison of exponential autocorrelation times for $q=10$ and\\
        & the algorithms SW, H, MH, and DQ on set $q=2,\ldots,10$.\\
\end{tabular}

\end{center}


\small\normalsize

\begin{thebibliography}{99}
\bibitem{sw}R.H. Swendsen and J.-S. Wang, Phys. Rev. Lett. {\bf 58}, 86 (1987).
\bibitem{bn}B.A. Berg and T. Neuhaus, Phys. Lett. {\bf B267}, 249 (1991).
\bibitem{bn2}B.A. Berg and T. Neuhaus, Phys. Rev. Lett. {\bf 68}, 9 (1992).
\bibitem{fs}A.M. Ferrenberg and R.H. Swendsen, Phys. Rev. Lett. {\bf} 61, 2635
(1988).
\bibitem{bip}A. Billoire, R. Lacaze, A. Morel, S. Gupta, A. Irb\"ack, and
B. Petersson, Nucl. Phys. {\bf B358}, 231 (1991).
\bibitem{ja1}W. Janke, HLRZ Preprint 50/92, J\"ulich (1992).
\bibitem{mp}E. Marinari and G. Parisi, Europhys. Lett. {\bf 19}, 451 (1992).
\bibitem{po}R.B. Potts, Proc. Camb. Phil. Soc. {\bf 48}, 106 (1952); for a
review see F.Y. Wu, Rev. Mod. Phys. {\bf 54}, 235 (1982).
\bibitem{bil}A. Billoire, R. Lacaze, and A. Morel, Nucl. Phys. {\bf B370}, 773
(1992).
\bibitem{jbk}W. Janke, B.A. Berg, and M. Katoot, Nucl. Phys. {\bf B382}, 649
(1992).
\bibitem{ru}R. Rummukainen, CERN Preprint CERN-TH.6654/92 (1992).
\bibitem{ls}X.-J. Li and A.D. Sokal, Phys. Rev. Lett. {\bf 63}, 827 (1989).
\bibitem{ke}W. Kerler, Nucl. Phys. B (Proc. Suppl.) {\bf 26}, 626 (1992);
Phys. Rev. {\bf D 47}, R1285 (1993).
\bibitem{add} It is to be noted that for MH the autocorrelation time does not
enter the error in the usual way. Comparing on the basis of variances, MH
looses
an additional factor of up to 2.3 in the range considered, which is {\it not}
included in the numbers given.
\end{thebibliography}
\end{document}